# Nanoparticle-Protein Interaction: Demystifying the Correlation Between Protein Corona and Aggregation Phenomena

*Larissa Fernanda Ferreira,[1,2#] Agustín Silvio Picco,[3#] Flávia Elisa Galdino,[1,4] Lindomar Jose Calumby Albuquerque,[1] Jean-François Berret,[5*] Mateus Borba Cardoso[1,2,4*]*

[1] *Brazilian Synchrotron Light Laboratory (LNLS) and Brazilian Nanotechnology Laboratory (LNNano), Brazilian Center for Research in Energy and Materials (CNPEM), Zip Code 13083-970, Campinas, Brazil.*
[2] *Programa de Pós-Graduação em Biotecnociências, Universidade Federal do ABC, Zip Code 09210-580, Santo André, Brasil.*
[3] *Instituto de Investigaciones Fisicoquímicas Teóricas y Aplicadas (INIFTA), Fac. de Cs. Exactas, Universidad Nacional de La Plata - CONICET, Boulevard 113 y 64, Zip Code 1900, La Plata, Argentina*
[4] *Institute of Chemistry (IQ), University of Campinas (UNICAMP), P.O. Box 6154, Zip Code 13083-970, Campinas, Brazil*
[5] *Matière et Systèmes Complexes, UMR 7057 CNRS Université Denis Diderot Paris-VII, Bâtiment Condorcet, 10 rue Alice Domon et Léonie Duquet, 75205 Paris, France.*

\* Corresponding authors
J.F.B.: jean-francois.berret@u-paris.fr
M.B.C.: cardosomb@lnls.br
[#] These authors equally contributed to this work

### Abstract

Protein corona formation and nanoparticles' aggregation have been heavily discussed over the last years since the lack of fine-mapping of these two combined effects has hindered the targeted delivery evolution and the personalized nanomedicine development. We present a multi-technique approach that combines Dynamic Light and Small-Angle X-ray Scattering techniques with cryo-Transmission Electron Microscopy in a given fashion that efficiently distinguishes protein corona from aggregates formation. This methodology was tested using ~ 25-nm model silica nanoparticles incubated with either model proteins or biologically relevant proteomes (such as fetal bovine serum and human plasma) in buffers of low and high ionic strengths to precisely tune particle-to-protein interactions. In this work, we were able to differentiate protein corona, small aggregates formation, and massive aggregation, as well as obtain fractal information of the aggregates reliably and straightforwardly. The strategy presented here can be expanded to other particle-to-protein mixtures and might be employed as a quality control platform for samples that undergo biological tests.

### 1. Introduction

The lack of fundamental understanding of physicochemical processes that occur when nanoparticles are injected into the bloodstream has hampered meaningful advances in nanoparticles-based precision medicine.[1–3] Inside the body, nanoparticles (NPs) non-specifically adsorb blood proteins (forming the so-called protein corona) which end up masking the outermost surface of the engineered NPs and represent one of the main mistargeting reasons.[4–7] Furthermore, the protein corona is prone to induce NPs aggregation, limiting their ultimate application since the overall surface-area is drastically changed, reducing their biological outcome. Also, the surface charge and composition (dictated by the protein corona) as well as the size and shape of particles or aggregates are intimately correlated to undesirable coagulation disorders and immune responses.[7–9]

Within this scenario, many works have been devoted to deeply investigate the formation of protein corona onto nanoparticles,[9–16] and how these nanoparticle-protein complexes interact with cells and trigger biological responses.[17–20] However, the colossal complexity behind these studies dealing simultaneously with NPs, protein coronas, and biological systems has made them rely on model systems where concentrations and conditions are far from those envisaged in real bloodstream applications. This strategy has definitively boosted an exponential advance on "quasi-ideal" protein corona understanding at the same time that these findings have found significant limitations in being correlated with medical reality resulting in low bench-to-market translation.[18,21] In addition, the desire for obtaining ideal systems for protein corona studies has somehow prevented an in-depth understanding of the ultimate role of this protein shell on triggering or preventing NPs aggregation in biological fluids.[8,12,22] This fact associated with the difficulty of distinguishing between protein corona and NPs aggregation has prevented a broader understanding and the possibility of establishing any correlation between these two phenomena. Also, the techniques and strategies currently available are usually more suitable for studying either protein corona or NPs aggregation.[12,23,24] Although many reports have tried to draw a correlation between protein corona and NPs aggregation, the amount of information is still sparse, and the correlation between them is not obvious.[22,25,26] In parallel, mimicking NPs in natural medical environments is challenging,[7] and only a very few and low informative number of techniques can finely probe these structures when in blood.[23,27–29] Further, the complexity related to the ionic strength of the media and the vast span in concentration ranges of all components are bottlenecks that limit consistent and rational nanoparticle-based medical development.[8,12,30–34]

Here we report an innovative experimental design that spans over an extensive protein-to-particle concentration range and provides an entire picture that considers protein corona formation and NPs partial or massive aggregation states close to natural medical conditions. Silica nanoparticles of ~ 25 nm were first synthesized and later incubated in buffers at distinct ionic strengths with either bovine albumin serum, bovine γ-globulin, chicken egg lysozyme, or fetal bovine serum. These complex mixtures of nanoparticles and proteins were then in-depth investigated by applying a cutting-edge association of complementary Dynamic Light and Small-Angle X-ray Scattering techniques (DLS and SAXS, respectively) with Cryo-Transmission Electron Microscopy (Cryo-TEM). The approach described here was able to finely map protein corona formation and its NPs aggregation induction in a concentration-dependent fashion as required for a rational understanding of sustainable and effective systems for medicine use.

## 2. Materials and Methods

### 2.1. Materials

Tetraethoxysilane (TEOS, reagent grade 98%), ammonium hydroxide aqueous solution (NH$_4$OH: 28-30%.wt NH$_3$, ACS reagent), sodium phosphate monobasic (NaH$_2$PO$_4$), sodium phosphate dibasic (Na$_2$HPO$_4$), phosphate buffer saline (PBS) tablets, cellulose membrane (cut-off 12-14 kDa), bovine serum albumin (BSA), lysozyme and bovine γ-globulin (BGG) were purchased from Sigma-Aldrich (St. Louis, MO, USA). Ethanol (P.A.) was obtained from Merck (Darmstadt, HE, Germany). Fetal bovine serum (FBS) was purchased from Gibco (part of ThermoFisher

Scientific, Whatham, USA; Cat. Num.: 12657-029, Lot Num.: 210420K, protein composition informed by the suppleier: total proteins = 49.3 g.L$^{-1}$, albumin = 27.2 g.L$^{-1}$, α-globulins 19.1 g.L$^{-1}$, β-globulins 3 g.L$^{-1}$ and γ-globulins 0.119 g.L$^{-1}$). Water used in all procedures was obtained from a water purification system (Purelab from ELGA – BUX, UK) and had a measured resistivity of 18.2 MΩ·cm. All buffers and protein solutions were filtered (pore 0.22 µm) prior to use.

### 2.2. Synthesis and characterization of silica nanoparticles (SiO$_2$)

SiO$_2$ synthesis was adapted from procedures already described in the literature.[35–37] In summary, 3.5 mL of NH$_4$OH was added to 120 mL of ethanol under magnetic stirring. Two aliquots of 2.5 mL of tetraethyl orthosilicate (TEOS) were then added to the pre-formed mixture after 0.5 and 3 h of stirring. Stirring was continued for another 24 h at room temperature and the formed SiO$_2$ were purified by dialysis using a cellulose membrane and washed first in an ethanol/water mixture of (75:25 v/v) and then in water until neutral pH. The resulting aqueous suspension was stored at 4 °C.

SiO$_2$ concentration in the aqueous suspension was determined by thermogravimetric analysis (TGA) using a thermogravimetric analyzer Pyris 1 (Perkin-Elmer). During this procedure, 30 µL of SiO$_2$ suspension were heated from 20 to 200 °C under a synthetic air atmosphere with a heating rate of 5 °C per minute. After evaporation, sample weight attains a plateau above 110-120 °C which is attributed to the remaining solid content (SiO$_2$). The value of the remaining weight (at 150 °C) was used to calculate the concentration of SiO$_2$ in the aqueous suspension. The analysis was performed in quintuplicate.

SiO$_2$ hydrodynamic diameter ($D_H$) was evaluated by dynamic light scattering (DLS) using a Malvern Zetasizer ZS equipment equipped with a red laser (632.8 nm) and operated in backscatter mode (detection angle = 173°). The measurements were performed in triplicate, where each measurement consisted in 10 runs of 10 seconds, at 25°C with thermal stabilization of 120 seconds. The correlation curves were analyzed using the method of cumulants to obtain the hydrodynamic diameter (Z-average) and the polydispersity index (PDI), and a non-negative least squares adjustment algorithm (NNLS) to extract size distributions. Both procedures are implemented in the Malvern´s Zetasizer software.

SiO$_2$ zeta potential measurements were also done using the Malvern Zetasizer ZS equipment above-mentioned. SiO$_2$ suspensions (1 mg.mL$^{-1}$) were prepared in a phosphate buffer (PB, prepared from NaH$_2$PO$_4$ and Na$_2$HPO$_4$) at pH = 7.4 and sonicated prior to the experiment. Measurements were made in triplicates, at 25 °C (with 120 seconds of thermal stabilization), using an applied potential of 150 V.[38]

The shape, (dry) size and size distribution of nanoparticles were evaluated by transmission electron microscopy (TEM). The samples were prepared by depositing a drop of the SiO$_2$ suspension on a 400 mesh microscope slide coated with a 5 nm carbon film. Samples were examined using a JEOL 2100 microscope (0.25 nm point-to-point resolution) with an acceleration voltage of 200 kV. The experiments were carried out at the Brazilian Nanotechnology Laboratory (LNNano) electron microscopy facility. Data processing was performed using ImageJ software.[39] In order to obtain size distribution at least 300 nanoparticles were counted .

### 2.3. Nanoparticle-Protein Mixtures Preparation

Four sources of distinct model proteins were used throughout this study: bovine serum albumin (BSA), bovine γ-globulin (BGG), chicken egg lysozyme and fetal bovine serum (FBS). The interaction between these proteins and SiO$_2$ was evaluated in either 10 mM phosphate buffer (PB, prepared from NaH$_2$PO$_4$ and Na$_2$HPO$_4$) or phosphate buffered saline (PBS, from tablets containing 10 mM PB + 137 mM NaCl + 2.7 mM KCl) both at pH 7.4. The BSA, BGG and lysozyme solutions were prepared by solubilizing the proteins in PB or PBS. After filtration (0.22 µm), the protein concentration was determined by UV-Vis absorption (λ = 280 nm) using the absorption coefficients 0.67 L.g$^{-1}$ cm$^{-1}$, 1.38 L.g$^{-1}$ cm$^{-1}$ and 2.51 L.g$^{-1}$ cm$^{-1}$ for BSA, BGG and lysozyme, respectively.[40] Subsequently, appropriate dilutions were performed to achieve the working concentrations. The FBS solutions were prepared by dilution in a volumetric flask since the total protein concentration (49.3 g.L$^{-1}$) is reported by the manufacturer.

Nanoparticle-protein mixtures were prepared by mixing 2 g.L$^{-1}$ SiO$_2$ suspensions and 2 g.L$^{-1}$ of the chosen protein solutions in PB or PBS and keeping the final volume of the system and the total mixture concentration in g.L$^{-1}$ constant. Thus, each mixture can be described by its mixing ratio (volume of protein solution: volume of SiO$_2$ suspension; the mixing ratio is equivalent to the ratio between protein and SiO$_2$ mass concentrations). In addition, each mixture can be described by its protein-to-particle molar ratios, PPMR. In this case, for each mixture, protein molarities were calculated considering their molecular weights and SiO$_2$ molarity was calculated taking into account nanoparticle size (by TEM) and mass concentration. More details on the calculations can be found in the results and discussion section and in the supporting information (SI; see **section S1**). Human plasma samples were obtained from the blood of healthy individuals from the Unicamp blood center (according to institutional bioethics approval – CAEE: 45209821.7.1001.5404), pooled in aliquots and stored in a freezer before the experiments. Total protein plasma concentration was determined with a bicinchoninic acid (BCA) assay according to the manufacturer's protocol (Thermo). Nanoparticle-protein mixtures were studied by dynamic light scattering (DLS), small X-ray angle scattering (SAXS) and cryogenic transmission electron microscopy (cryo-TEM). All experiments were performed within 30 min after mixture preparation. Inside this time frame, we could not detect any structural change in the studied systems. DLS experiment were performed using the same equipment explained above. SAXS measurements were used to determine the size and polydispersity of SiO$_2$ and proteins, as well as to obtain structural information regarding protein corona or aggregates formation upon the mixtures. The measurements were performed on the D1B-SAXS1 and Cateretê beamlines in the Brazilian Synchrotron Light (LNLS) facility. Silver behenate was used as a standard to calibrate the instrument and. All measurements were performed at room temperature and, prior to data processing, all the corresponding corrections were made on the two-dimensional image obtained during the experiment. Modeling of SAXS data was performed using SASfit software.[41]

Cryo-TEM was used to investigate protein-induced SiO$_2$ aggregation. Briefly, 7µl of the suspension were dripped onto the grid and the excess was removed with a filter paper. The grid was then immersed in liquid ethanol and liquid nitrogen (Vitrobot FEI) and transferred to the TEM JEOL 1400 PLUS microscope. The analyzes were performed at a voltage of 120 kV and the experiments were carried out at the electron microscopy facility of LNNano.

### 3. Results and Discussion

In this work, silica nanoparticles (SiO$_2$) were obtained by the sol-gel method through a one-step synthesis approach[35,36,37] and purified by dialysis. TEM was used to investigate the shape, size, and size distribution of the synthesized particles. **Figures 1a** and **b** show selected images where SiO$_2$ of uniform sizes and quasi-spherical shapes can be seen at low and high magnifications, respectively. A Gaussian size distribution (**Figure 1c**) was used to adjust the TEM data, and a mean diameter of D$_{TEM}$ = 23.2 nm, and standard deviation δ = 2.6 nm were obtained. In addition, DLS measurements were performed to evaluate SiO$_2$ hydrodynamic diameter (D$_H$) and polydispersity (PDI). A D$_H$ of 30.7 nm and a considerably low PDI of 0.089 were observed by cumulant analysis in phosphate buffer (PB). **Figure 1d** shows the corresponding SiO$_2$ intensity-weighted D$_H$ distribution obtained by the NNLS method. Similar results were also obtained in saline conditions (See SI, **section S2**). The differences between the average sizes obtained by TEM and DLS can be ascribed to the fundamentals behind each technique and have been well documented in the literature.[42–44] The zeta potential of the particles was also measured in 10 mM phosphate buffer at pH = 7.4, and a value of -39.9 ± 2.5 mV was obtained, which is an indication of the dissociated silanol groups on the SiO$_2$ surface and agrees well with previously reported values.[45,46]

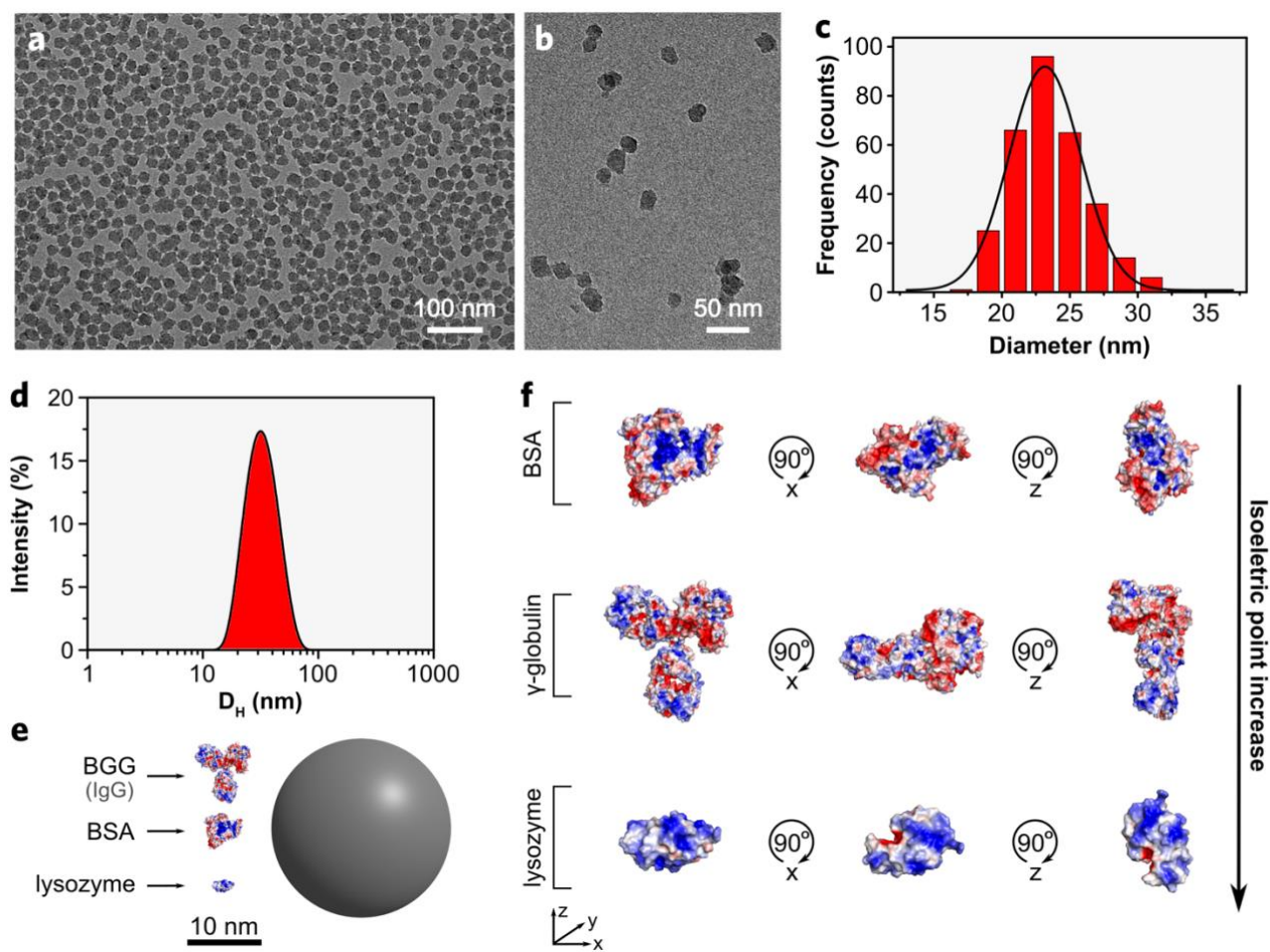

*Figure 1. Structural details of silica nanoparticles and proteins used throughout this work. (**a**) Low and (**b**) high-magnification transmission electron microscopy images of silica nanoparticles. Silica nanoparticle size distributions derived from (**c**) transmission electron microscopy images and (**d**) dynamic light scattering-based SiO$_2$ size distribution (**e**) Size comparison between silica nanoparticle and bovine γ-globulin (BGG), bovine serum albumin (BSA), and lysozyme. Since IgGs are the main constituents of the BGG fraction, an IgG molecule was presented. (**f**) Electrostatic potential surfaces of BSA, BGG (IgG), and lysozyme. Positive regions are represented in blue, negative ones are shown in red, while white areas depict neutral regions. All protein structures were taken from the Protein Data Bank.[47] PDBs were 3v03 for BSA, 3WUN for lysozyme, and 1IGY for IgG. Molecular representations were made using PyMOL.[48] Electrostatic potential surfaces were calculated with APBS[49] (as implemented in PyMOL).*

These nanoparticles were incubated with either bovine serum albumin (BSA), bovine γ-globulin (BGG), chicken egg lysozyme, or fetal bovine serum (FBS) in phosphate buffer (PB, pH =

7.4, ionic strength, I ~ 25 mM, calculated from ion composition as I = ½ $\Sigma_i c_i z_i^2$ where $c_i$ and $z_i$ are the concentration and ionic valency of the i[th] ion in the media, respectively) or phosphate buffer saline (PBS, pH = 7.4, I ~ 163 mM). These two buffers were selected to evaluate the influence of ionic strength (low in PB and high in PBS). Serum albumins are the most abundant protein in mammals' sera (e.g., in humans, > 50 % of total protein blood content) and play a fundamental role as carriers (of drugs, fatty acids, ions, among others) as well as on the maintenance of oncotic pressure. BSA is a ~ 66.5 kDa molar mass (Mw) protein formed by a single polypeptide chain containing 583 amino acid residues, and its isoelectric point (pI) is between 4.70 and 5.60, giving rise to an overall negative charge at pH = 7.4. In addition, it has an overall structure akin to human serum albumin (HSA) due to their similarity in 76% of the amino acid sequence[50] and justifies the large number of studies with BSA. γ-globulins are an essential protein fraction of mammals' sera mainly constituted by immunoglobulins (antibodies), where the IgGs are the most abundant ones. Immunoglobulins are critical players of the humoral immune response.[51] BGG is a mixture of immunoglobulins (approximately 80% IgG, 10%, IgM and 10% IgA)[52] with an average molecular weight of ~ 150 kDa (as informed by the distributor) and a pI ~ 6.6 (thus, presenting a subtle negative charge at pH=7.4).[53] Lysozyme is a ubiquitous globular protein with bactericidal properties found in human fluids like saliva and tears.[54] Lysozyme from chicken egg (the most commonly used lysozyme source) has a polypeptide chain constituted by 129 amino acid residues and presents an Mw ~14 kDa.[55] It has a high pI ~ 11.4, which results in a positively and strongly charged structure at pH = 7.4. Lysozyme has been frequently used as a small and hard protein model, and independent studies have shown that its adsorption onto silica nanoparticles (negatively charged) is considerably strong due to electrostatic interactions.[56–58] Thus, lysozyme has been used as a model for analyzing the $SiO_2$ behavior in the positively charged protein presence even though its concentration in plasma is negligible. A size comparison (in scale) between a 25 nm silica particle and BSA, BGG (IgG), and lysozyme is presented in **Figure 1e**. It highlights the size difference between the particle and the proteins and points out a significant difference when these proteins are compared. An out-of-scale representation of proteins' electrostatic potential surfaces is shown in **Figure 1f**. Here, positive areas are represented in blue, negative domains are shown in red, while neutral regions are depicted in white. It is not obvious to identify a more significant number or more concentrated blue/red areas, although we have presented BSA, γ-globulin, and lysozyme from low to high pIs. Irrespective of their overall charge at physiological pH (BSA negative, BGG barely negative, and lysozyme positive), the three proteins have abundant positive and negative domains. Differently from the three formerly discussed protein sources, FBS is a complex mixture of proteins, lipids, sugars, free amino acids, hormones, vitamins, trace elements, among other components. In addition, FBS composition is batch-dependent [59] and presents the protein fraction as its primary component with concentrations ranging from 32 to 70 g.L$^{-1}$. Among this protein fraction, albumin is the principal component (e.g., in the FBS used here represents ca. 55%) while low to negligible concentrations of BGG is found. More details of the FBS used here are provided in the materials section. Relevant data on the different proteins or sources of proteins used along the work is collected in **Table 1**.

**Table 1.** Molecular weight, pI, and charge at physiological pH (7.4) of the proteins understudy

|  | Mw (KDa) | pI | Charge at pH=7.4 |
|---|---|---|---|
| **BSA** | 66.5 | 4.7-5.6 | (-) |
| **BGG** | 150 | 6.6 | Slightly (-) |
| **Lysozyme** | 14 | 11.4 | (+) |
| **FBS** | *83\** | - | - |

*This value is a tentative molecular weight calculated under the rough approximation that the proteins in FBS other than albumin can be represented as having an Mw ~104 KDa.[60] See the text below for more details.

The interaction between $SiO_2$ and the proteins mentioned above (BSA, BGG, Lysozyme, and FBS) was studied by varying the molar concentration of nanoparticles and proteins while keeping the total concentration (mass/volume) constant. This approach is based on the Job's Plot method (method of continuous variations),[61–63] commonly used in analytical chemistry to determine binding stoichiometry over large concentration ranges, and has been successfully adapted by our group to investigate polyelectrolytes association through DLS.[64–67] **Figure 2** shows the $D_H$ evolution at various protein-to-particle molar ratios (PPMR; see SI **section S1**) for distinct proteins and buffers. For FBS, since its protein fraction is a complex mixture, a tentative Mw was estimated by taking a weighted average between BSA and the rest of the proteins' content, considering that the remaining proteins can be roughly represented by an Mw ~ 104 KDa.[60] Thus, we calculated a value of Mw ~ 83 KDa for the FBS used here. This value was considered in all PPMR calculations involving FBS.

**Figure 2** shows the $D_H$ evolution of $SiO_2$ at various PPMRs for distinct proteins in PB and PBS. The interaction between $SiO_2$ and BSA in PB (**Figure 2a**) indicates a slight $D_H$ increase, suggesting BSA adsorption on the $SiO_2$ surface, forming a protein corona without aggregation. The maximum $\Delta D_H$ is around 3 nm, which is less than the 6-8 nm expected for a monolayer of BSA with protein molecules adsorbed on a flat configuration.[68] This is likely related to the free BSA concentration that increases with increasing PPMR, thus biasing the observed nanoparticle $D_H$ toward lower values. Similar trends have been already reported in the literature.[69] At higher ionic strength, in PBS (**Figure 2b**), BSA adsorption on $SiO_2$ induces nanoparticles aggregation, evidenced by the significant $D_H$ increase of about 10-fold (see **Figures 2c** and **c**). In this case, $SiO_2$-BSA complexes aggregate likely by screening the electrostatic interactions, resulting in a surface charge reduction and consequent mitigation of repulsion between the particles. This hypothesis is supported by considering that the number of proteins required to induce the maximum $D_H$ decreased from approximately 100 to 10 when going from PB to PBS.

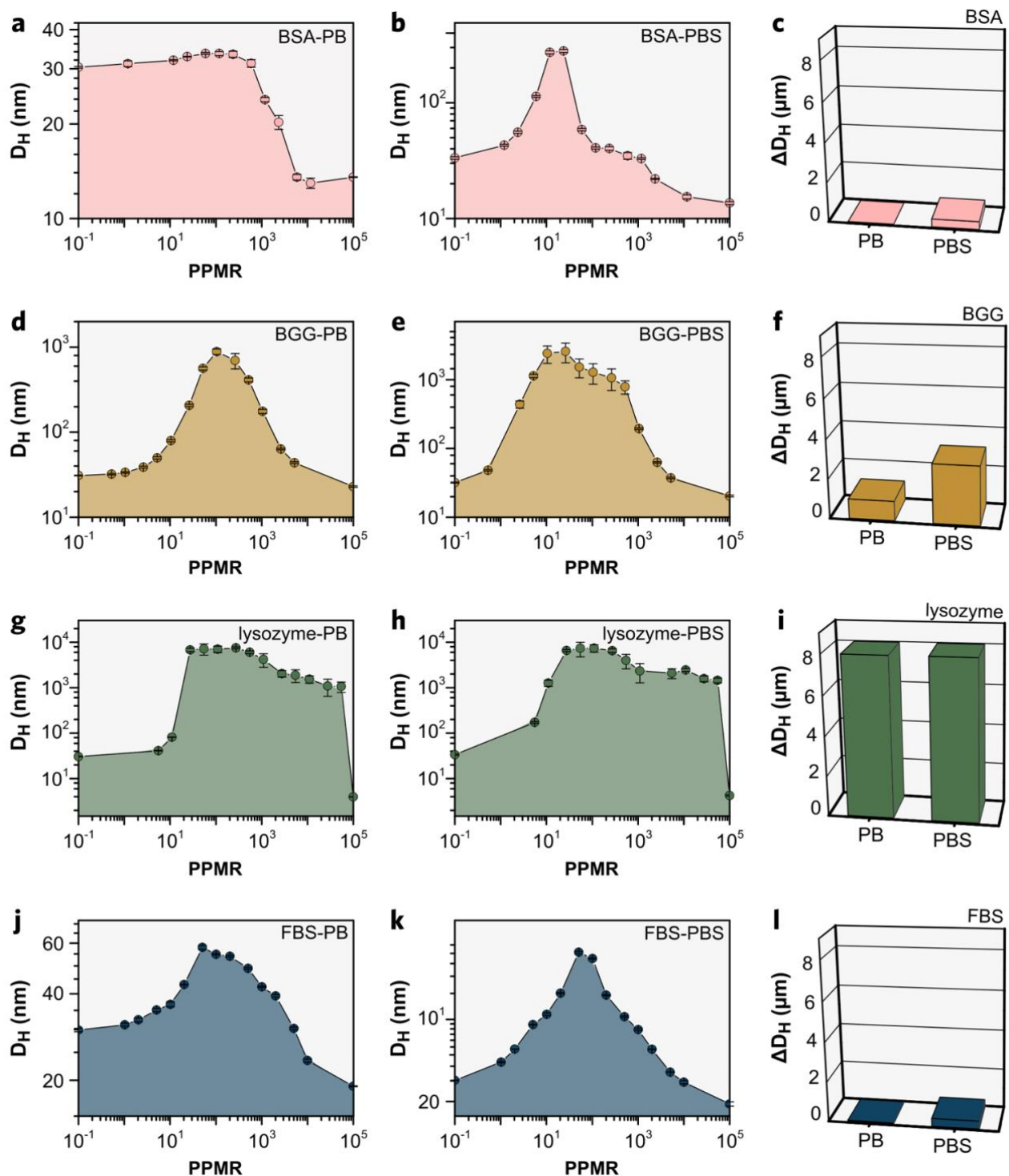

***Figure 2.*** *Hydrodynamic diameter ($D_H$) evolution as a function of protein-to-particle molar ratio (PPMR) in panels **a**, **b**, **d**, **e**, **g**, **h**, **j**, and **k** and maximum hydrodynamic diameter variation compared to the bare silica particle ($\Delta D_H$) in different media in panels **c**, **f**, **i** and **l**. Data for BSA in panels **a**, **b**, and **c**, for BGG in panels **d**, **e** and **f**, for lysozyme in panels **g**, **h** and **i** and FBS in panels **j**, **k**, **l**. The left-most and right-most experimental points in panels **a**, **b**, **d**, **e**, **g**, **h**, **j**, and **k** are silica in the absence of proteins and proteins in the absence of silica, respectively.*

For BGG, no signals of aggregation can be seen in PB at very low ratios, and we assume that protein corona is formed when PPMR is smaller than 10 (**Figure 2d**). However, a sudden $D_H$ increase can be observed, reaching a maximum aggregate size of about 1 µm at PPMR ~ 100. It likely indicates that the overall protein charge plays an important role, at least at high protein/particles ratios, since the reduction of the overall protein negative charge (BGG) is no longer enough to maintain the colloidal stability of the system as observed for BSA. This scenario is further confirmed by the data in PBS (**Figure 2e**), where aggregates much larger than the ones found for BSA in PBS are seen (**Figure 2f**). Coincidently, the PPMR values for the highest ΔD$_H$ are similar for BSA and BGG when the same buffer is taken into account. This likely indicates that the adsorption/aggregation is driven by the same mechanism, further enhanced when the protein pI is increased. Contrarily, lysozyme induced massive particles aggregation in both media (**Figure 2g** and **h**) from extremely low to high PPMRs. Final aggregate sizes are considerably larger than 5 µm (**Figure 2i**), and, consequently, we cannot accurately estimate their sizes nor discuss the ionic strength effect. In this specific case, the charge neutralization between the positively charged proteins and the negatively charged particles drives to a strong interaction, leading to large nanoparticle aggregates in both media. Furthermore, it has been previously reported that lysozyme can be the bridging point between distinct silica particles, further contributing to their aggregation.[56,70] As described above, FBS is a mixture of various proteins which in the presence of PB adsorbed on the SiO$_2$ surface without signals of massive aggregation (**Figure 2j**). In addition, the ΔD$_H$ maximum was close to 30 nm, which either indicates the formation of clusters composed of a few particles or a thick multilayered protein corona, as already reported in the literature.[71] **Figure 2k** shows the results for SiO$_2$-FBS mixtures in PBS that exhibits a maximum ΔD$_H$ of approximately 300 nm at PPRM around 50-100. Furthermore, the overall Job´s plot shape (by comparing **Figures 2a** and **j**, and **Figures 2b** and **k**) and the maximum ΔD$_H$ (**Figure 2c** and **l**) are reasonably similar to the one observed for BSA, indicating that the dominant behavior of FBS is dictated by albumin which is the most abundant component. Differences between the interaction of SiO$_2$ with FBS and BSA might be associated with other components found in FBS. It is essential to point out that D$_H$ tends to decrease with increasing PPMR after reaching the D$_H$ maxima gradually. This behavior can be attributed to the excess of free protein that biases the D$_H$ values toward lower values, as discussed for BSA in PB. Contrarily, lysozyme forms very large aggregates, and the scattering contribution coming from the free proteins is almost indistinguishable, while the D$_H$ values only marginally decrease at high PPMRs. These results are summarized in **Table 2**. In addition to these DLS results, the evolution of the total scattering intensity with increasing PPMR for all the different mixtures understudy was also recorded and exhibited similar trends observed in **Figure 2** (results presented in the SI - **Section S4**).

**Table 2.** Summary of main results derived from DLS and SAXS analysis

| | DLS | SAXS |
|---|---|---|

|  | PB | | PBS | | PB | | PBS | |
| --- | --- | --- | --- | --- | --- | --- | --- | --- |
|  | PPMR$_{max}$ | ΔD$_H$ | PPMR$_{max}$ | ΔD$_H$ | PPMR$_{max}$ | P$_{max}$ | PPMR$_{max}$ | P$_{max}$ |
| **BSA** | 50-200 | 3 | 10-25 | 250 | - | 0 | 200 | 2.5 |
| **BGG** | 100 | 850 | 10-25 | 3000 | 25-1000 | 3-4 | 5-500 | 4 |
| **Lysozyme** | 25-550 | > 5000 | 25-300 | > 5000 | > 5 | 2-3-2.7 | 10-300 | 2.3-2.5 |
| **FBS** | 50 | 30 | 50-100 | 550 | 2-200 | < 2.3 | 20-50 | 3.5 |

*ΔD$_H$ and P$_{max}$ refer to the maximum hydrodynamic diameter obtained by DLS and the maximum P-value adjusted by SAXS, respectively. PPMR$_{max}$ refers to the PPMR value at which ΔD$_H$ or P$_{max}$ are observed.

Complementarily, SAXS was employed as a tool to differentiate protein corona and aggregation as well as to obtain structural information related to the aggregates. **Figure 3a** shows representative SAXS curves resulting from the mixture between SiO$_2$ and lysozyme in PB. The SAXS curves for the remaining samples are presented in the SI (see **Section S6**).

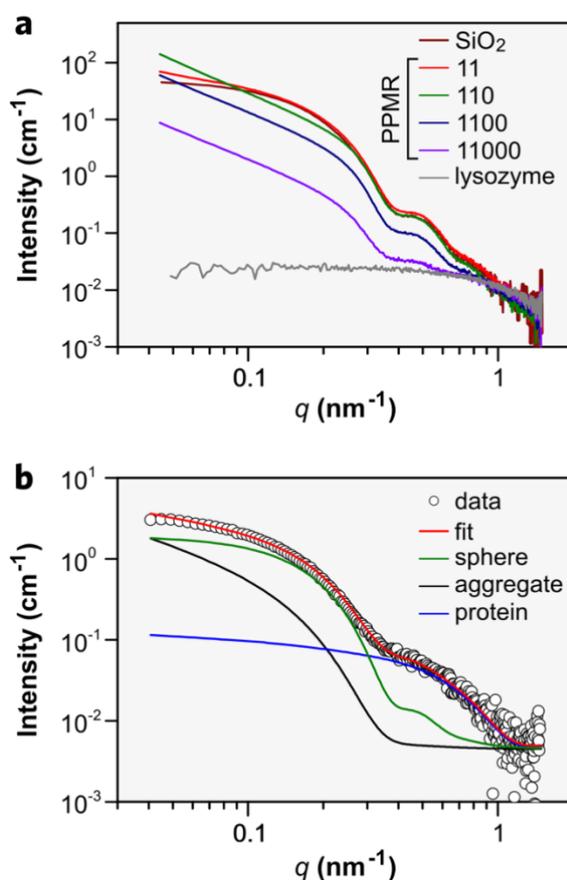

*Figure 3. (a) Small-angle X-ray scattering (SAXS) profiles for pure lysozyme and SiO$_2$ as well as for a few selected mixtures between these two components at distinct protein-to-particle molar ratios (PPMR) in PB media. (b) An experimental SAXS curve of a sample containing BSA and SiO$_2$ at PPMR = 20 in PB was hypothetically chosen to highlight the structural levels used for fitting the data. The red curve corresponds to the global fit, the blue level signal comes from BSA, the green level is attributed to silica nanoparticles, while the black level intensity is related to the aggregate signal.*

Meaningful SAXS profile changes can be seen and are related to the changes in sample composition or the formation of aggregates. In the low-$q$ part ($q < 0.2$ nm$^{-1}$), the scattering is dominated by the presence of individual structures (usually nanoparticles at low PPMRs or proteins at high PPMRs) or the formation of aggregates which induce an evident change of the scattering curve (mainly, the inclination in the low-$q$ region). On the other hand, for $q$ values larger than 0.2 nm$^{-1}$, the scattering is mainly dominated by SiO$_2$, proteins, or their physical mixture (we could not detect cooperative interaction). Then, the SAXS curves were deconvoluted by using up to three different scattering contributions: (a) SiO$_2$ nanoparticles (represented by polydisperse spheres), (b) proteins and, (c) aggregates fitted using a Power-law decay (P) together with a cut-off constrained to the size of SiO$_2$.[72] Details on the used model can be found in the SI (see **Section S4**). A deconvolution example is presented in **Figure 3b**, where all three distinct contributions can be identified. In all cases, SiO$_2$ particles and proteins were first fit using the scattering of pure systems, and upon the mixtures, only scale factors were used, which drastically reduced the number of free parameters.

P values were then obtained for each mixture, and the results are presented as Job´s plots in **Figure 4**. For BSA in PB, P ~ 0 implies the absence of aggregation, suggests that only protein corona formation occurs, and corroborates DLS results. For this specific case, subtle changes in $D_H$ (DLS) could not qualitatively affect the SAXS profiles, and all curves overlap when scaling is applied. Except for FBS in PB, all remaining samples have shown maximum P values either around 2.5 or between 3 and 4, meaning either mass or surface fractal formation, respectively.[73] Maximum P values deriving from the interaction of SiO$_2$ and the different protein sources under study are summarized in **Figure 4 c**, **f**, **i**, and **l**. Although we cannot extract aggregates size from SAXS curves since a shoulder-type Guinier region is not seen in the low-$q$ region,[72–75] the P values are helpful to distinguish fractal features for the distinct aggregates. Non-equilibrium growth processes likely related to diffusion-limited aggregation are usually related to P ~ 2.5, suggesting that this kind of structure is not closely packed. This behavior is observed when SiO$_2$ interacts with BSA in PBS (**Figure 4b**) and with lysozyme in both media (**Figure 4g** and **h**). On the other hand, much denser structures are foreseen for surface fractals where P values are between 3 and 4. This is observed for the interaction of SiO$_2$ with BGG, both in PB and PBS (**Figure 4d** and **e)** and for FBS in PBS (**Figure 4k**). Contrarily to the scenario described above, SiO$_2$ exposed to FBS in PB has shown P-values around 2. Although it is possible to argue that this could also be related to non-equilibrium growth processes that produce less dense aggregates, such as multiparticle diffusion-limited aggregation, other possibilities are also sound.[73] Among them, it would also be possible to consider that these small structures induce an intermediate P-value between 0 and 2.3 or 3.0-4.0 and, consequently, do not reach the final aggregate state. In addition, the possibility of forming small clusters of a few particles with irregular shapes would also give rise to a P-value which could also be close to 2. **Table 2** presents the main results obtained by both SAXS, while comparing them with the DLS ones.

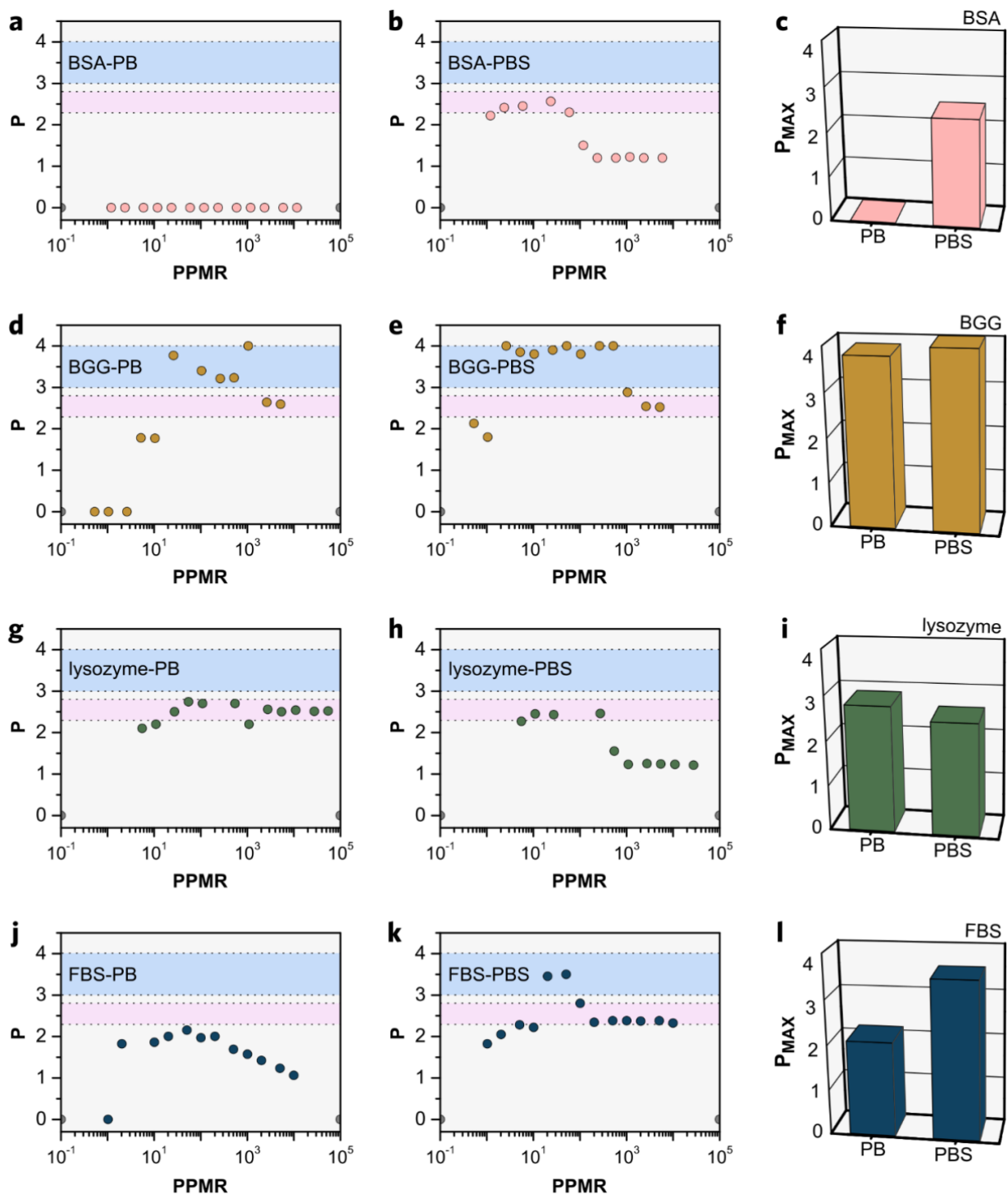

*Figure 4.* Power-law decay (P) evolution as a function of protein-to-particle molar ratio (PPMR) in panels *a*, *b*, *d*, *e*, *g*, *h*, *j*, and *k* and the maximum P-value of each panel ($P_{MAX}$) in different media in panels *c*, *f*, *i*, and *l*. Data for BSA in panels *a*, *b*, and *c*, for BGG in panels *d*, *e*, and *f*, for lysozyme in panels *g*, *h*, and *i*, and for FBS in panels *j*, *k*, and *l*. Blue and pink regions correspond to P values between 3 and 4 and 2.3 and 2.8, respectively. The left-most and right-most experimental points in panels *a*, *b*, *d*, *e*, *g*, *h*, *j*, and *k* are silica in the absence of proteins and proteins in the absence of silica, respectively.

Cryo-TEM was then used to elucidate the correlation between size and structure in the four SAXS scenarios described above (P ~ 0: no aggregation; P ~ 2: small clusters or weakly dense aggregates; P ~ 2.5: mass fractals; P > 3: surface fractals). Representative images are shown in **Figure 5**, while additional micrographs are provided in the Supporting Information. **Figure 5a** shows the mixture of $SiO_2$ and BSA in PB at PPMR ~ 25, which corresponds to P = 0, where individual particles are seen. In this case, we understand that protein corona formation occurs due to the slight $D_H$ increase with no qualitative change in SAXS curves. In addition, this image highlights that the changes seen by DLS are not related to a partial aggregation as schematized in **Figure 5b**.

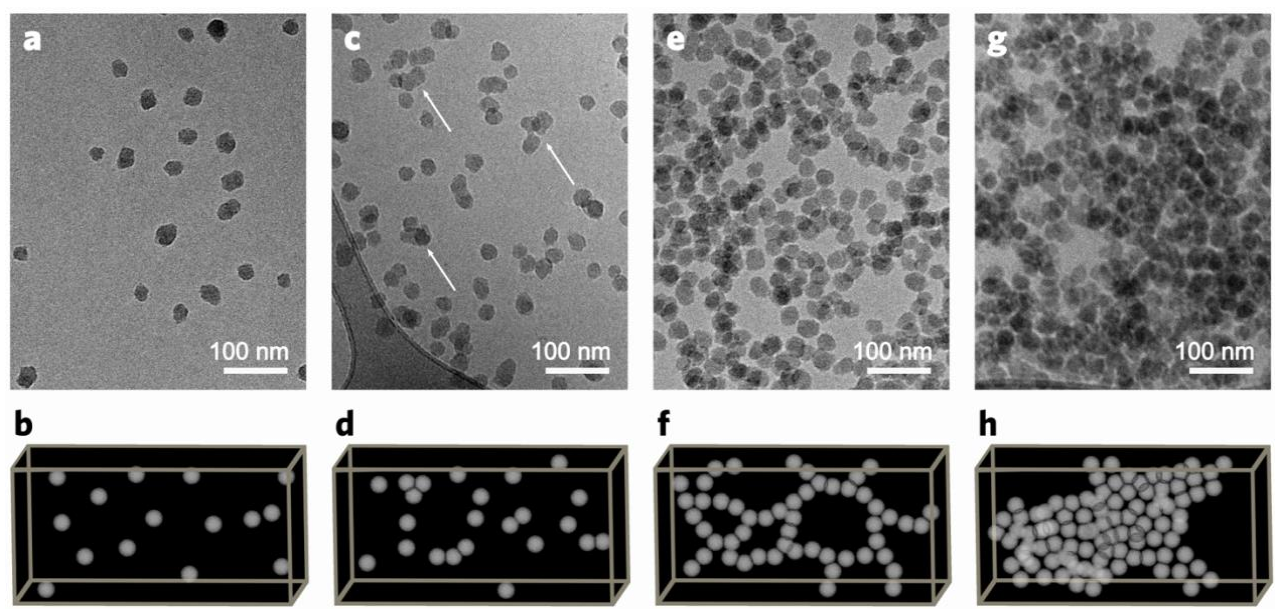

*Figure 5. Cryo-transmission electron microscopy images (**a, c, e** and **g**) and schematic structural representations (**b, d, f** and **h**) for selected samples presenting P = 0 (**a** and **b**), P ~ 2.0 (**c** and **d**), P ~ 2.5 (**e** and **f**) and 3 > P > 4 (**g** and **h**). (**a**) $SiO_2$ and BSA in PB at PPMR ~ 25, (**c**) FBS in PB at mixing ratio = 0.5, (**e**) lysozyme in PB at PPMR ~ 270 and (**g**) BGG in PBS at PPMR ~ 25. These images clearly show that the larger the P values induce more compact structures.*

**Figure 5c** shows a cryo-TEM image for $SiO_2$ and FBS in PB at PPMR ~ 45, corresponding to P ~ 2. A quasi-homogeneous sample is seen in this specific case where a few small aggregates (indicated by the arrows) are identified. These aggregates are likely the reason for a significant $D_H$ increase from 30 to 85 nm and are also correlated to the power-law appearance in the SAXS curves. Taking this image into account, we can now ascribe the found P-value (~ 2) for SAXS patterns to the inhomogeneity of the sample, which is a convolution resulting from distinct form factors. Thus, the presence of small and inhomogeneous aggregates (as schematized in **Figure 5d**) induces a SAXS profile change at low-$q$. The combination of the characterization tools used here clearly points out that the $D_H$ increase observed by DLS cannot be uniquely attributed to the protein corona since these small aggregates induce a $D_H$ increase and the appearance of P ~ 2. **Figures 5e** and **g** show two representative cryo-TEM images for aggregated samples representing structures with SAXS curves showing P ~ 2.5 (mixture of $SiO_2$ and lysozyme in PB at PPMR ~ 270) and between 3 and 4 (mixture of $SiO_2$ and BGG in PBS at PPMR ~ 25), respectively. Individual particles are rarely seen, which indicates massive aggregation in both cases. Although the DLS technique cannot discriminate them, SAXS P values can indicate about their differences, which is now confirmed by cryo-TEM images. The interaction between SiO$_2$ and lysozyme produces micrometric structures, which are seen by DLS, while the SAXS P ~ 2.5 indicates a mass fractal structure that agrees well with an open structure (**Figure 5e**). Consequently, aggregates resembling an open network are observed, and **Figure 5f** shows a schematic representation of this aggregation process. Contrarily, a much denser structure is observed for cryo-TEM images (**Figures 5g**) obtained from the mixtures of SiO$_2$ and BGG, where the SAXS P-values are between 3 and 4. Furthermore, the diminished amount of holes and the reduced void volume inside the aggregate structure corroborate the surface fractal nature previously suggested by the SAXS analysis (i.e., 3 < P < 4). **Figure 5h** presents a schematic representation of the surface fractal aggregates presenting a denser, more compact, and holes-free structure.

As a proof-of-concept of our approach applicability, silica nanoparticles were then incubated with platelet-free human plasma. Human plasma was obtained through full blood centrifugation and, consequently, presents all its coagulation components. Therefore, it is a very heterogenous sample in terms of size and composition and makes data interpretation very challenging. Before the experiment, human plasma protein concentration was determined and this value was used as quantification point for the mixtures and the graphs were then plotted as a function of protein-to-particle mass ratio. **Figure 6** shows the $D_H$ evolution of SiO$_2$ when incubated with human plasma in PB and PBS. The overall shape of the patterns is similar to the ones presented in **Figure 2**. Protein corona formation cannot be ascribed or detected here since even a physical mixture without interaction will result in an overall $D_H$ increase. This increase is due to the large size measured for human plasma when compared to bare silica which will result in a size increase that might be related to either protein corona formation or simply the presence of structures that are larger the. Nevertheless, we observe a significant $D_H$ increase for those intermediate mass ratios irrespective of the media. This $D_H$ increase is attributed to the formation of large aggregates constituted by SiO$_2$ and plasma proteins (or other constituents). Moreover, considering that human plasma is full of coagulant factors, that the anticoagulant is diluted in the particle-protein mixtures (the lower the mass ratio, the higher the dilution), and that silica is a known coagulant activator,[76,77] partial coagulation might take place in the mixtures.

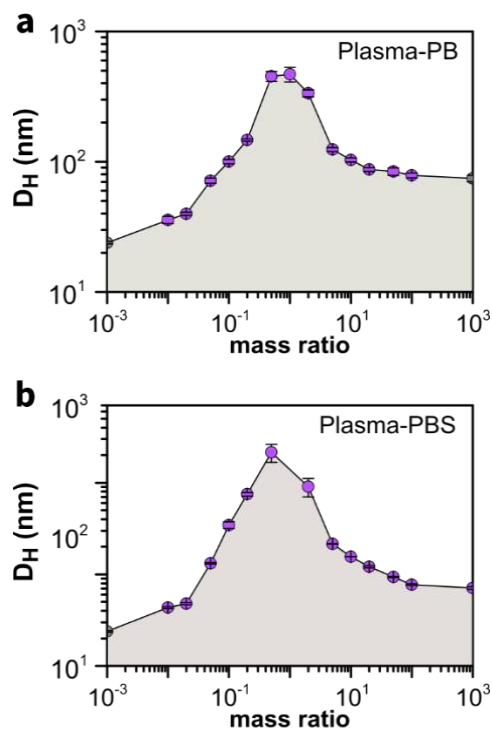

***Figure 6.*** *Hydrodynamic diameter ($D_H$) evolution of mixtures containing silica nanoparticles and human plasma as a function of protein-to-particle mass ratio in PB (a) and PBS (b). The left-most and the right-most experimental points in panels refers to silica in the absence of proteins and proteins in the absence of silica. The left-most and right-most experimental points in both panels are silica in the absence of proteins and proteins in the absence of silica, respectively.*

Then, SAXS measurements were performed to give any clue about the nature of these large structures seen by DLS (**Figure 6**). The SAXS curves (**Figure SX**) were exclusively fit using linear combinations of $SiO_2$ nanoparticles and human plasma scattering contributions. For these specific set of samples, differently from **Figures 3** and **4**, the third scattering level ($I_{agg}$) related to the nanoparticles aggregate formation was not needed to reach a good fitting quality (**Section S4 – SI**). However, the nature of the mixture here is very different and allow us to infer a few things. The bare human plasma present large structures which are seen by DLS (**Figure 6**). The scattering intensity is dependent on the size of the object and on the electron density difference between the mixture components. Consequently, while the human plasma present some large structures, silica nanoparticles present higher electron density. Thus, these large structures from the human plasma could, hypothetically, be masking the nanoparticles aggregation signal due to the large scattering power attributed to their size. However, if this was the case, we would probably identify a meaningful scattering profile change at low-*q* that would agree with the large DLS variation and with the scattering patterns in the presence of model proteins (**Figure S4**). Thus, although we observe a very important DLS change upon the mixtures, we believe it cannot be attributed to the particle aggregation due to the absence of $I_{agg}$ for the $SiO_2$ nanoparticles and human plasma mixtures. Alternatively, we hypothesize that the DLS variation followed by the absence of SAXS $I_{agg}$ can be ascribed to the partial plasma coagulation. We suggest that coagulation takes place likely induced by the presence of the $SiO_2$. It likely means that $SiO_2$ is triggering the aggregation of biological components which are not directly seen in the SAXS curves due to either their low scattering contrast (hidden by the silica scattering power) or their size that generates scattering changes out of our experimental window (at lower *q* values).

The presented strategy can be expanded and adjusted to other particle-protein mixtures and employed as a quality control platform before biological tests. To further illustrate this point and highlight the relevance of the proposed strategy, we will extrapolate our approach to a well-established U. S. Food and Drug Administration (FDA) or European Medicine Agency (EMA) approved nanomedicine. We first consider a patient weighing 70 kg and height of 170 cm (surface body area, SBA ~1.8 $m^2$).[78] Thus, the patient total blood volume is ~5 L while 50% hematocrit (meaning ~2.5 L of total plasma volume) and regular plasma values for BSA (~45 $g.L^{-1}$, ~$6.8·10^{-4}$ M) and γ-globulin (~11.5 $g.L^{-1}$, ~$6.7·10^{-5}$ M) are considered.[79] DOXIL®, the first-approved and one of the most established anticancer nanomedicines in the market,[80] consists of PEGylated liposomes (liposome concentration of ~$4.5·10^{16}$ $particles.L^{-1}$, or ~$7.5·10^{-8}$ M) carrying doxorubicin (at a suspension concentration of 2 $g.L^{-1}$)[81] and has been used to treat ovarian cancer (dose of 50 mg of doxorubicin per $m^2$ of SBA), Kaposi´s sarcoma (20 $mg.m^{-2}$) and myeloma multiple (30 $mg.m^{-2}$). For instance, ovarian cancer treatment requires 45 mL of intravenously DOXIL administration into the patient described above. Thus, the DOXIL liposome plasma concentration is ~$1.3·10^{-9}$ M and results in PPMRs of ~$5·10^5$ and $5·10^4$ for BSA and γ-globulins, respectively. Consequently, the strategy described here can precisely map individual interactions between proteins and nanoparticles and can be used to probe the overall stability of the nano-system towards major and critical blood components taking into account their natural abundance. Moreover, it is essential to foresee that our strategy can also be brought to a three-dimensional view where nanoparticles and more than a single component can be simultaneously investigated, expanding our finds relevance in terms of practical use.

## 4. Conclusions

The use of a multi-technique approach combining Dynamic Light and Small-Angle X-ray Scattering techniques with cryo-Transmission Electron Microscopy to characterize and efficiently distinguish protein corona from aggregates formation was successfully demonstrated. The experiments were done with 25 nm silica particles in the presence of either model proteins (bovine albumin serum, bovine γ-globulin, and chicken egg lysozyme) or biologically relevant media (such as fetal bovine serum proteome or human plasma) in buffers of low and high ionic strengths. The particle-to-protein interaction was precisely tunned employing the Job's Plot method (continuous variation method) to cover extended concentration ranges and mimic environmental conditions relevant to intravenous nanoparticles' administration. The multi-technique approach proposed here was able to discriminate between protein corona and small aggregates formation as well as massive nanoparticles' aggregation. Furthermore, this approach was sensitive to distinguish between different types of aggregates and, consequently, extract their fractal information. We suggest that the strategy presented here can be employed to investigate complex particle-to-protein systems in conditions similar to those found during *in vivo* tests.

## 5. Acknowledgements

The authors acknowledge the financial support of the Fundação de Amparo à Pesquisa do Estado de São Paulo (FAPESP – processes 2019/24894-7, 2017/02145-7, 2018/09555-9, and 2015/25406-5). M.B.C. acknowledge CNPq for a productivity fellowship. A.S.P acknowledge FAPESP for a former fellowship (grant number 2014/21910-8). A.S.P. is staff member of CONICET. The authors acknowledge Electron Microscopy Laboratory of LNNano for the use of electron microscopy facility (proposal TEM-24820), and LNLS for the use of SAXS1 (SAXS1-20180366) and Cateretê (SAXS1-20180366) beamlines. We are very thankful to Florian Meneau, Aline Passos, Tiago Kalile and Paulo Ricardo Garcia for the valuable help during SAXS acquisition and data reduction done at Cateretê beamline.

**Graphical Abstract**

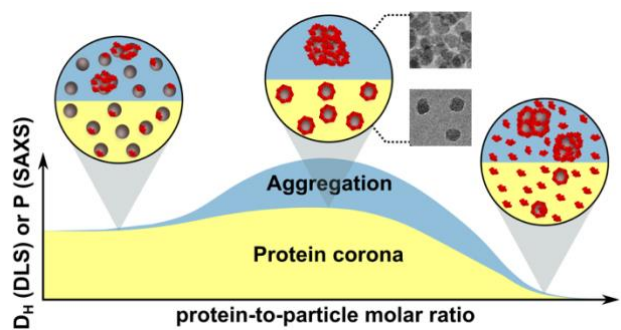